\documentclass[aps,prb,twocolumn, groupedaddress]{revtex4-1}

\usepackage{graphicx}
\begin{document}


\title{Single domain to multi-domain transition due to in-plane magnetic anisotropy in phase separated (La$_{0.4}$Pr$_{0.6}$)$_{0.67}$Ca$_{0.33}$MnO$_{3}$ thin films}


\author{Hyoungjeen Jeen}
\author{Amlan Biswas}
\email{amlan@phys.ufl.edu}
\affiliation{Department of Physics, University of Florida, Gainesville, Florida 32611, USA}


\date{\today}

\begin{abstract}
  Phase separated perovskite manganites have competing phases with different crystal structures, magnetic and electronic properties. Hence, strain effects play a critical role in determining the magnetic properties of manganite thin films. Here we report the effect of anisotropic stress on the magnetic properties of the phase separated manganite (La$_{0.4}$Pr$_{0.6}$)$_{0.67}$Ca$_{0.33}$MnO$_{3}$. Thin films of (La$_{0.4}$Pr$_{0.6}$)$_{0.67}$Ca$_{0.33}$MnO$_{3}$ grown under anisotropic in-plane stress on (110) NdGaO$_{3}$ substrates display in-plane mangetic anisotropy and single domain to multidomain transition as a function of temperature. Angle dependent magnetization measurements also show that the magnetization reversal occurs mainly through the nucleation $\&$ propagation mechanism. By comparing the results with (La$_{0.4}$Pr$_{0.6}$)$_{0.67}$Ca$_{0.33}$MnO$_{3}$ thin films grown on (001) SrLaGaO$_{4}$ substrates, we have confirmed that the magnetic anisotropy is mainly due to substrate induced anisotropic stress. Our results suggest novel avenues for storing magnetic information in nanoscale magnetic media.  
\end{abstract}


\maketitle

\section{Introduction}
    The coupling between structure, transport, and magnetism in hole-doped manganites leads to phenomena such as, colossal magnetoresistance (CMR), colossal electroresistance (CER), photo-induced metal-insulator transition, and colossal piezoresistance (CPR)~\cite{RefWorks:92, Dha07, Miy97, Asa97, Tos09}. While these properties could lead to future applications in devices such as bolometers and cryogenic memories, manganites are already providing a unique insight into the effect of competing phases on the physical properties of materials~\cite{RefWorks:133, RefWorks:116}. It is now widely accepted
that phenomena such as CMR are consequences of the competition among different phases with similar free energy. Such competition leads to phase coexistence among three distinct phases, \textit{viz.} cubic ferromagnetic metallic (FMM), 
pseudo-tetragonal (more precisely orthorhombic) antiferromagnetic charge ordered insulating (AFM-COI), and pseudo-cubic paramagnetic insulating (PMI) phases, in materials such as (La$_{1-y}$Pr$_{y}$)$_{1-x}$Ca$_{x}$MnO$_{3}$ ~\cite{RefWorks:83, RefWorks:99, RefWorks:105}. In addition to well-known effects such as CMR and CER, the coexistence of the three magnetic phases also leads to unique phenomena such as temperature dependent magnetic domain transition and ellipsoidal growth of the FMM phase, which have been observed using Lorentz microscopy in very narrow temperature ranges~\cite{RefWorks:96}.
Due to the same coupling between crystal structure, transport, and magnetism, manganite thin films have shown properties distinct from bulk behavior such as substrate strain induced metal-insulator transition and anisotropic transport due to strain fields from substrates~\cite{RefWorks:84, RefWorks:101}. The effect of strain on the transport properties of manganites has been widely studied and it is accepted that multiphase coexistence and percolation play a significant role~\cite{RefWorks:99, RefWorks:105, Dha07}. However, the effect of strain on the magnetism of phase separated manganites is more subtle and is still being debated. One such problem is the distinction between intrinsic magnetic properties and extrinsic effects on magnetic properties of manganite thin films. For example, when La$_{0.77}$Ca$_{0.33}$MnO$_{3}$ thin films were grown on single crystalline NdGaO$_{3}$ substrates, Mathur \textsl{et al.} concluded that the in-plane magnetic anisotropy originated not from stress anisotropy but from magnetocrystalline anisotropy~\cite{RefWorks:94}.     
Here, we report that substrate induced stress plays an integral role in determining the magnetic properties of manganite thin films. We observe that in-plane stress anisotropy leads to an in-plane magnetic anisotropy and a magnetic domain transition as a function of \textit{temperature} in phase separated (La$_{1-y}$Pr$_{y}$)$_{1-x}$Ca$_{x}$MnO$_{3}$  ($x$ = 0.33 and $y$ = 0.6) thin films grown on (110) NdGaO$_{3}$ substrates with anisotropic in-plane strain. Our data show that while anisotropic stress has a profound effect on the magnetism, the in-plane resistivity of the films remains virtually isotropic. By comparing our results for the films on anisotropic NGO to those grown on isotropic (001) SrLaGaO$_{4}$ (SLGO) substrates, we conclude that anisotropic strain can be used to control the magnetic ``hardness" i.e. the coercive field in a mixed phase manganite. Such control could play an important role in the design of nanomagnetic devices. 

\section{Experimental details}
    (La$_{1-y}$Pr$_{y}$)$_{1-x}$Ca$_{x}$MnO$_{3}$ (LPCMO, $x$ = 0.33 and $y$ = 0.6) thin films of two different thicknesses (30 nm and 20 nm) were grown on orthorhombic (110) NdGaO$_{3}$ (NGO) substrates by pulsed laser deposition (PLD) (KrF, $\lambda$ = 248 nm). The substrate temperature during growth was 780$^{\circ}$C, O$_{2}$ partial pressure was 130 mTorr, laser energy density was about 0.5 J/cm$^{2}$, and growth rate was kept at about 0.4 \AA/s. Step flow growth has been consistently observed under these conditions in the LPCMO thin films up to 60 nm thickness. The magnetic response reported in this article was observed in four different films. The thickness was controlled by deposition time and then confirmed by atomic force microscopy~\cite{afm}. The lattice mismatch strains in the two in-plane directions of LPCMO thin films grown on NGO substrates (LPCMO//NGO) are $\delta_{1\bar{1}0} = 0.49\%$ and $\delta_{001} = 0.26\%$ due to different in-plane lattice parameters of the NGO substrates ($d_{1\bar{1}0}$ = 3.863 \AA, $d_{001}$ = 3.854 \AA, $d_{LPCMO}$ = 3.844 \AA)~\cite{RefWorks:109, Vas00}, where $\delta (\%) \equiv \frac{d_{substrate}-d_{film}}{d_{film}} \times 100$. We also grew 30-nm-thick LPCMO thin films on (001) tetragonal SrLaGaO$_{4}$ (SLGO) substrates (LPCMO//SLGO) using the same growth parameters given above. Since the substrate induced strain on LPCMO//SLGO thin films is negligible due to well matched in-plane lattice parameters of SLGO ($d$ = 3.842 \AA), the films grown on the two different substrates can be used to isolate the effect of anisotropic strain on the magnetism and transport of LPCMO. The structure of the films was characterized by conventional $\theta-2\theta$ x-ray diffraction using a Philips APD 3720 diffractometer. 
Magnetic properties were measured using a Quantum Design 5 T Superconducting Quantum Interference Device (SQUID) magnetometer. Since LPCMO thin films show thermal hysteresis in both magnetic and transport properties, thermal demagnetization was executed by heating up to 150 K (a temperature higher than the hysteretic region and the magnetic Curie temperature, $T_C \approx 130$ K) before each measurement. Field demagnetization of the superconducting magnet was also performed at 150 K. For measuring the magnetic moment as a function of temperature ($M(T)$), a 100 Oe field was applied to minimize magnetic field induced phase change of the LPCMO films~\cite{RefWorks:125}. Two different methods were used to remove the background paramagnetic signal from the NGO substrates and obtain the magnetic moments of LPCMO films. The first method was direct subtraction for which, magnetic moment as a function of applied field ($M(H)$) measurements were carried out at different temperatures for the bare NGO substrates before film deposition. $M(T)$ was also measured for the same substrates in a field of 100 Oe. After deposition of the LPCMO films on the substrates, $M(H)$ and $M(T)$ curves were acquired under the same conditions as the background measurements. The $M(H)$ and $M(T)$ curves of the substrates were then subtracted from the corresponding $M(H)$ and $M(T)$ curves of the film plus substrate to obtain the magnetic moments of the LPCMO thin films. The second method was based on linear background fitting. The raw $M(H)$ data of LPCMO films have signals both from the  ferromagnetic films and the paramagnetic NGO substrates. Since the paramagnetic signal is linear at low fields, the background signal can be obtained by fitting a linear function to the data from 700 Oe to 2000 Oe (i.e. magnetic fields greater than the coercive field), and then subtracted from the raw data to obtain the magnetic moments of LPCMO films. The linear background fitting procedure can be used only for the $M(H)$ measurements. Background subtraction was not required for LPCMO films grown on non-magnetic SLGO substrates. A Digital Instruments Multimode scanning probe microscope was used in the tapping atomic force microscope mode to check the surface morphology and thickness of the thin films~\cite{afm}. Resistivity as a function of temperature ($\rho(T)$) was measured in a helium cryostat equipped with a Janis variable temperature insert.

\section{Results and Discussion}
\subsection{Structure and Transport}
\begin{figure}[b]
\includegraphics[width=8.5cm,height=12cm] {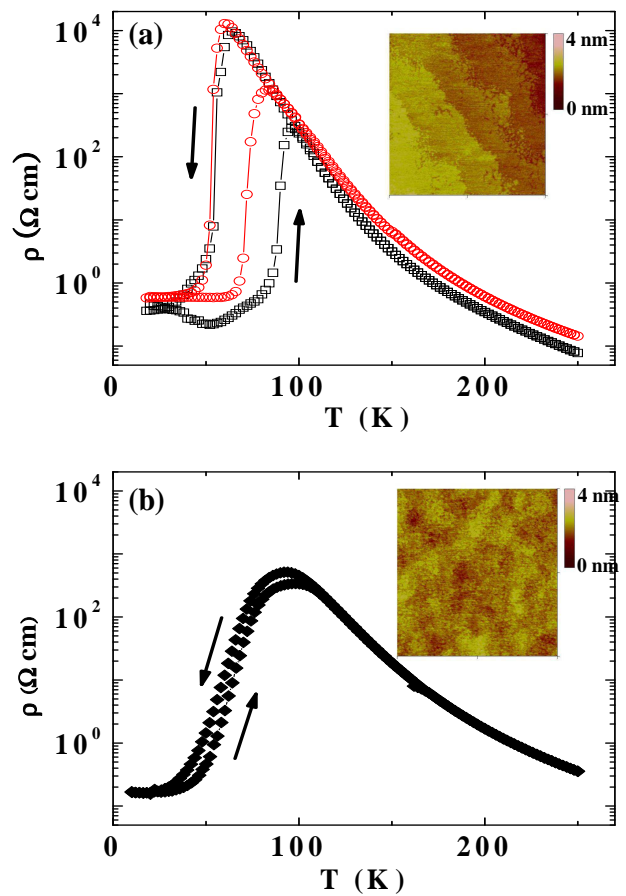}
\caption{(Color online) (a) Resistivity as a function of temperature ($\rho(T)$) of a 30-nm-thick (circle) and 20-nm-thick (square) (La$_{0.4}$Pr$_{0.6}$)$_{0.67}$Ca$_{0.33}$MnO$_{3}$ (LPCMO) thin films grown on (110) NdGaO$_{3}$ (NGO) substrates. The inset shows a 2 $\mu$m $\times$ 2 $\mu$m atomic force microscope (AFM) image of a 20-nm-thick LPCMO thin film grown on NGO. (b) $\rho(T)$ behavior of a 30-nm-thick LPCMO thin film grown on a SrLaGaO$_{4}$ (SLGO) substrate. The inset shows a 2 $\mu$m $\times$ 2 $\mu$m AFM image of a 30-nm-thick LPCMO thin film grown on SLGO.}
\end{figure}
    The surfaces of our LPCMO thin films on both NGO and SLGO substrates are smooth on an atomic scale (insets of Fig. 1(a) and 1(b)). The r.m.s. roughness of the LPCMO//NGO films is about 2 \AA. The films on NGO usually show step-flow growth mode with unit cell step heights as shown for the 20 nm-thick film (inset of Fig. 1(a)). The step height is about 4 \AA, which is comparable to the lattice constant of bulk LPCMO. The r.m.s. roughness of a 30-nm-thick LPCMO//SLGO film is about 4 \AA~ and step-flow growth was not observed. Resistivity measurements of the LPCMO//NGO films show sharp transitions at the insulator-metal transition temperature ($T_{IM}$, obtained while cooling) and metal-insulator transition temperature ($T_{MI}$, obtained while warming) whereas, the LPCMO//SLGO thin films  show a more gradual transition and narrower thermal hysteresis (Fig. 1(b))~\cite{tim}. The $T_{MI}$ of the 20-nm-thick LPCMO//NGO thin film was about 92 K, which is 16 K higher than that of the 30-nm-thick film, while the $T_{IM}$ are within 4 K (60 K for the 20-nm-thick film and 56 K for the 30-nm-thick film). The large variation of $T_{MI}$ with thickness is related to the strain-induced static phase separated (SPS) state at low temperatures and we are currently performing experiments to study this effect in detail~\cite{Dha07}. $\theta-2\theta$ x-ray diffraction pattern of the 30-nm-thick LPCMO//NGO film did not show any individual LPCMO peaks due to the similar lattice parameters of LPCMO and (110) NGO in the out-of-plane direction, while the x-ray diffraction pattern of the LPCMO//SLGO film clearly shows sharp LPCMO peaks (Fig. 2). Thus, all the LPCMO films were grown with a single chemical phase and were highly ordered along perpendicular direction to the substrate surface. The films are smooth on an atomic scale and show sharp resistivity transitions at $T_{IM}$ and $T_{MI}$.
\begin{figure}[t]
\includegraphics [width=8.5cm,height=6cm] {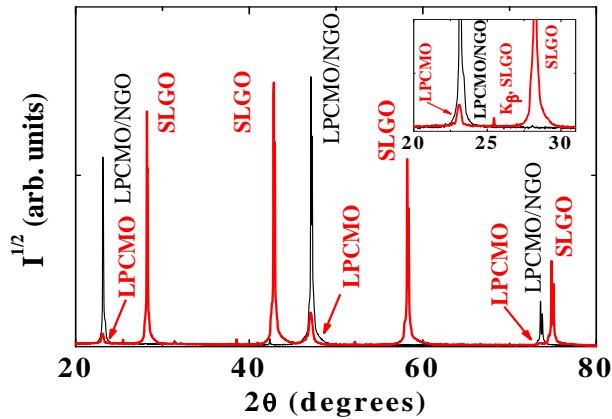}
\caption{(Color online) $\theta-2 \theta$ x-ray diffraction patterns of the 30-nm-thick LPCMO thin film grown on an SLGO substrate (thick line) and the 30-nm-thick LPCMO thin film grown on an NGO substrate (thin line). The inset shows the low angle region in detail.}
\end{figure} 

\subsection{Magnetic properties}
\begin{figure}[t]
\includegraphics [width=8.5cm,height=6cm] {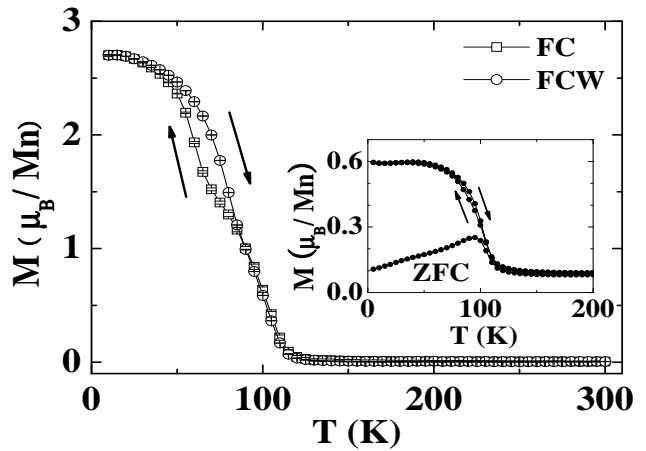}
\caption{Magnetic moment as a function of temperature ($M(T)$) of the 30-nm-thick LPCMO thin film grown on NGO under field cooling (FC) (square) and field cooled warming (FCW) (circle) runs in a 100 Oe field along the [1$\bar{1}$0] NGO direction. The inset shows $M(T)$ behavior of the 30-nm-thick LPCMO film gorwn on SLGO under zero field cooling (ZFC), field cooling (FC), and field cooled warming (FCW) in an in-plane 100 Oe field.}
\end{figure}
     Fig. 3 shows $M(T)$ curves of a 30 nm-thick LPCMO//NGO thin film, taken under field cooling (FC) and field cooled warming (FCW) in a field of 100 Oe applied parallel to the [1$\bar{1}$0] NGO direction. We will show in subsection D that the [1$\bar{1}$0] NGO direction is the easy axis for the LPCMO//NGO films. We used the direct subtraction method to get the pure magnetic moments of the LPCMO//NGO films. Due to the background subtraction, the zero field cooled (ZFC) $M(T)$ data is accomapnied by a large relative error and hence is not shown here. The $M(T)$ behavior is similar to previous results obtained for bulk LPCMO ($x$ = 0.375 and $y$ =0.600)~\cite{RefWorks:87}. The paramagnetic to ferromagnetic transition occurs at approximately 130 K. Although the entire film does not become ferromagnetic at 130 K since these films show multiphase coexistence, we define the $T_{C}$ to be approximately 130 K, at which $M$ becomes a measureable non-zero value~\cite{zfc}. While this definition of $T_C$ is not robust, it approximately marks the temperature at which the ferromagnetic regions are first nucleated. The $M(T)$ graph shows thermal hysteresis in the FC and FCW runs at similar temperatures where we observed hysteresis in the $\rho(T)$ behavior. This hysteresis is due to the fluid phase separated (FPS) state transforming into the glassy static phase separated (SPS) state at low temperatures~\cite{Dha07,RefWorks:125}. 
The magnetic moment saturates below 30 K as the LPCMO//NGO film enters the SPS region. In the SPS state the FMM regions are frozen in space and hence the magnetic moment is constant with a reduction in temperature ~\cite{Dha07,RefWorks:125}. The transition from the FPS to SPS state also leads to a unique behavior of the coercive field as a function of temperature to be described in the next section. $M(T)$ behavior of the LPCMO//SLGO film is shown in the inset of Fig. 3, which shows magnetic thermal hysteresis and saturation of the magnetic moment (0.6 $\mu_{B}/Mn$) below 30 K, similar to the LPCMO//NGO film.

\subsection{Variation of coercive field with temperature}
\begin{figure}[b]
\includegraphics [width=8.5cm,height=6cm] {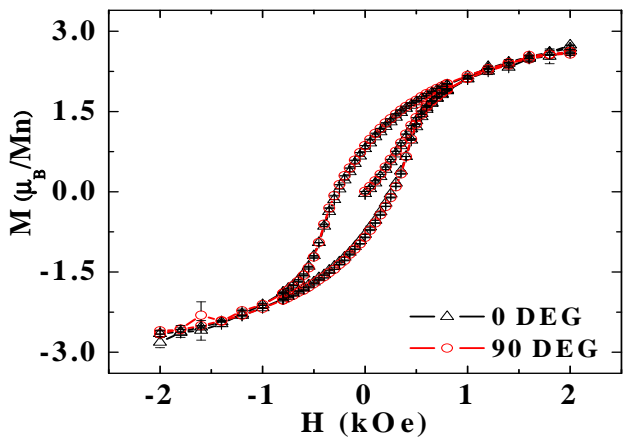}
\caption{(Color online) In-plane zero field cooled (ZFC) magnetization hysteresis loops of the 30-nm-thick LPCMO thin film grown on SLGO along the two orthogonal in-plane directions at 50 K.}
\end{figure}
     We will now discuss the effect of anisotropic strain on the magnetic properties of phase separated manganites. Previous studies have shown that in-plane anisotropic strain leads to uniaxial magnetic anisotropy in the plane of the thin film~\cite{RefWorks:108, RefWorks:94}. However, these studies were performed on purely ferromagnetic manganites. In a phase separated manganite, we expect that the magnetic anisotropy may lead to anomalous behavior of the submicrometer sized ferromagnetic regions~\cite{RefWorks:105}. Magnetic hysteresis loops of the thin films were measured in the zero field cooled (ZFC) state to study the pure unmagnetized state at each temperature. The magnetic moments of LPCMO//SLGO films could be directly measured, since SLGO substrate is non-magnetic. NGO substrates are paramagnetic, which necessitates a careful background subtraction to obtain the magnetization of LPCMO//NGO films. The $M(H)$ data were measured along two in-plane directions of  the SLGO substrate, \textit{viz.} [100] and [010]. Magnetic hysteresis loops in the two directions show that there is negligible magnetic anisotropy at all measured temperatures in terms of remanences and coercive fields as shown in Fig. 4 for $T=50$ K. The significance of these isotropic hysteresis loops will be discussed in subsection D. The $M(H)$ loops also show that the remanences are much lower than magnetic moments at 2 kOe (Fig. 4), which suggests weak interaction among ferromagnetic regions in the LPCMO//SLGO films~\cite{Cul08}. Due to the large paramagnetic signals from the NGO substrates we used two different methods \textit{viz.}, direct subtraction and linear background fitting to get ZFC magnetic hysteresis loops of the LPCMO//NGO films. The magnetic field is applied parallel to the [1$\bar{1}$0] direction of the NGO substrates which is also the magnetic easy axis of the films on NGO (subsection D). As shown in Fig. 5, the $M(H)$ loops obtained using direct subtraction and linear background fitting show similar results. For example, at 50 K direct subtraction gives a coercive field of 230 $\pm$ 4 Oe and a remanence of 2.4 $\pm$ 0.1 $\mu_{B}$/Mn while linear background fitting gives a coercive field of 230 $\pm$ 5 Oe and a remanence of 2.3 $\pm$ 0.1 $\mu_{B}/$Mn. The two background subtraction methods show similar results throughout the temperature range investigated. Compared to the LPCMO//SLGO films, the LPCMO//NGO films show rectangular $M(H)$ loops i.e. their saturation magnetization is comparable to their remanence. The sharp change of the magnetic moment at the coercive field suggests domain wall motion by either nucleation model or pinning model, and the rectangular shape of hysteresis loops implies a strongly developed uniaxial anisotropy in the LPCMO thin films on NGO substrates, as discussed in subsection D~\cite{Cul08}. 
\begin{figure}[t]
\includegraphics [width=8.5cm,height=12cm] {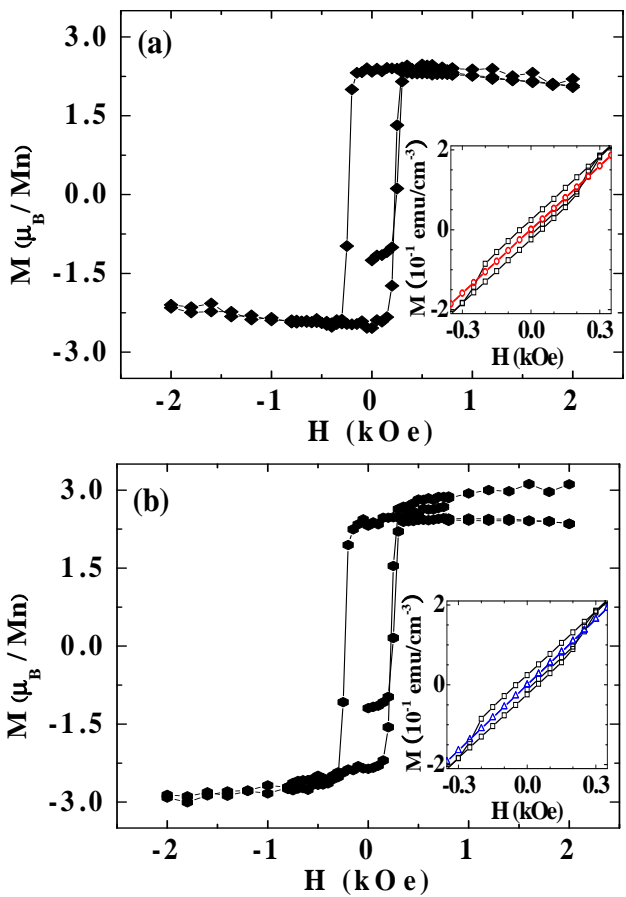}
\caption{(Color online) (a) ZFC magnetization hysteresis loop of the 30-nm-thick LPCMO thin film grown on NGO, at 50 K with the field applied along [1$\bar{1}$0] direction of the NGO substrate after using direct subtraction (DS). The inset shows $M(H)$ behavior of the substrate before deposition (circle) and the film plus substrate after deposition (square). (b) ZFC magnetization hysteresis loop of the 30-nm-thick LPCMO thin film grown on NGO, at 50 K with the field applied along [1$\bar{1}$0] direction of the NGO substrate after using linear background fitting. The inset shows $M(H)$ behavior of LPCMO and NGO (square) and the fitted paramagnetic background (triangle).}
\end{figure}  
        
   The $T_{C}$ of the LPCMO//NGO film is approximately 130 K (Fig. 3). From 130 K down to 80 K, the observed coercive fields are less than the field step size. While such small $H_{c}$ could be due to small single domain (SD) FMM regions in a matrix of PMI and/or COI phases, we believe that the remnant field in the superconducting magnet in a SQUID magnetometer gives rise to the observed hysteresis in this temperature range~\cite{qdusa}. Hence, we assume that $H_{c}$ $\approx$ 0 and that the film is in a super-paramagnetic state from 130 K down to 80 K. Below 80 K, the $H_{c}$ first increases sharply to about 300 Oe at 60 K and then decreases gradually to about 190 Oe at 30 K (Fig. 6(a)). However, $M_{r}$ increases monotonically as the temperature is decreased from 80 K to 30 K (inset, Fig. 6(a)). The behavior of $H_{c}$ and $M_{r}$ as a function of temperature shows that the FMM regions grow in the same temperature range in which the coercive field shows non-monotonic behavior. The growth of the FMM regions with lowering temperature is supported by magnetic force microscopy, Lorentz microscopy, transport measurements, and other magnetic measurements~\cite{RefWorks:105, RefWorks:99, RefWorks:96}. The non-monotonic behavior of $H_{c}$ as a function of temperature along with the monotonic increase of $M_{r}$ strongly suggests an SD to multidomain (MD) transition as a function of temperature. SD FMM regions are nucleated below $T_{C}$ and grow down to 60 K. Further growth of the FMM regions below 60 K results in MD behavior and hence, a decrease in $H_{c}$. Such magnetic domain transition is typically observed in ferromagnetic fine particles as a function of particle size~\cite{RefWorks:120}. Here we observe an SD to MD transition with decreasing temperature due to an increase in size of FMM regions embedded in a non-ferromagnetic AFM-COI/PMI matrix.      

    Below 30 K down to 5 K, the coercive field increases again (Fig. 6(a)). To explain this unexpected increase in $H_{c}$, we have to go back to Fig. 3, which shows that below 30 K LPCMO is in the SPS state~\cite{Dha07}. As a result, the size and distribution of the ferromagnetic regions remain constant below 30 K and the increase of coercive field may be due to the reduction of thermal energy in the multidomain ferromagnetic regions of constant size. To confirm this hypothesis we fit the $H_{c}(T)$ curve from 5 K to 30 K to the equation $H_{c}(T)/H_{c0}$ = $[1-(T/a)^{m}]$, where $a$ is related to the spin flip energy barrier at zero magnetic field, $H_{c0}$ is the coercive field at 0K, and $m$ is the exponent of temperature, since it has been observed that the coercive fields of multidomain nanoparticles show a $T^{2/3}$ dependence~\cite{Das10, Sko03}. From fitting, it is estimated that $m$ is 0.68 $\pm$ 0.02 ($H_{c0}$ $\approx$ 360 Oe and $a$ = $E_{0}/k_{B}$ $\approx$ 88 K) which confirms our hypothesis. The magnetic behavior of LPCMO in the low temperature region is also different from spin glass materials such as the CuMn system, where displaced hysteresis loops are observed after FC~\cite{Bin86}. When we measured magnetic hysteresis loops after FC to 10 K in a magnetic field of 10 kOe, we couldn't observe any significant difference in the positive and negative coercive fields (data not shown). When FC is carried out in a field of 10 kOe (which is higher than all the coercive fields measured at different temperatures), it increases the FMM phase of the thin film so that the domain transition sets in at the higher temperature of 70 K (Fig. 6(a)). In the LPCMO//SLGO films, all the magnetic quantities ($H_{c}$, $M_{r}$) show gradual increase as the temperature is reduced (Fig. 6(b)). In particular, the magnetic domain transition observed in LPCMO//NGO films was not observed in the LPCMO//SLGO films. We thus conclude that the domain transition is due to substrate induced anisotropic in-plane stress coupled with phase coexistence in LPCMO thin films. We now present a detailed description of the magnetic anisotropy in LPCMO thin films.  
\begin{figure}[t]
\includegraphics [width=8.5cm,height=12cm] {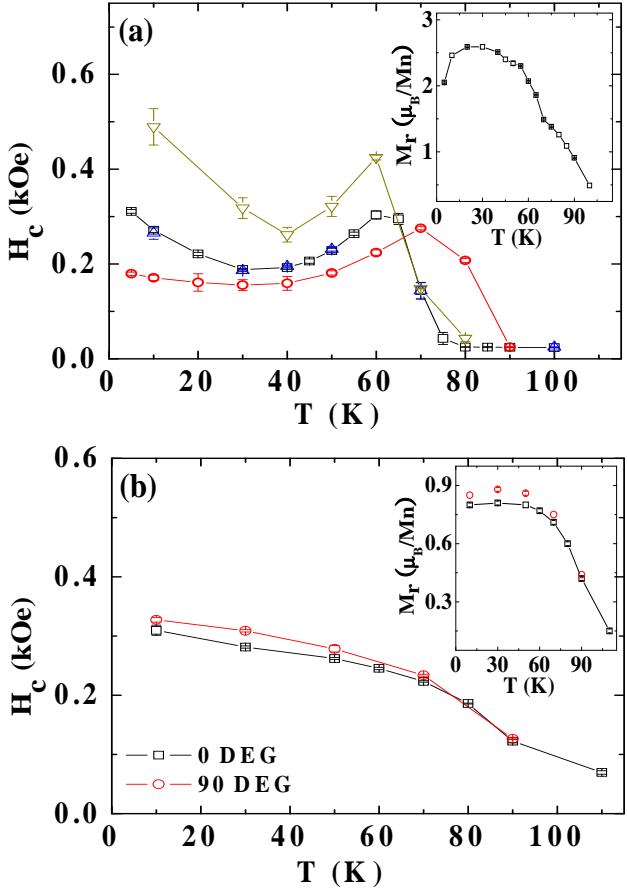}
\caption{(Color online) (a) Coercive field as a function of temperature ($H_{c}(T)$) for a 20-nm-thick LPCMO thin film grown on NGO from ZFC $M(H)$ curves with $H$ along [1$\bar{1}$0] NGO using linear background fitting (inverted triangle) and a 30-nm-thick LPCMO thin film grown on NGO from ZFC $M(H)$ curves with $H$ along [1$\bar{1}$0] NGO using linear background fitting (square) and direct subtraction (triangle) and 10 kOe FC $M(H)$ curves with field along [1$\bar{1}$0] NGO using linear background fitting (circle). The inset shows the remanence as a function of temperature ($M_{r}$(T)) for the 30-nm-thick LPCMO thin film grown on NGO from ZFC $M(H)$ curves with $H$ along [1$\bar{1}$0] NGO using linear background fitting. (b) $H_{c}(T)$ for a 30-nm-thick LPCMO film grown on SLGO from ZFC $M(H)$ curves with $H$ applied along  two orthogonal in-plane directions. The inset shows $M_{r}$(T) for the 30-nm-thick LPCMO thin film grown on SLGO from ZFC $M(H)$ curves with $H$ applied along two orthogonal in-plane directions.}
\end{figure}
\subsection{Strain induced magnetic anisotropy}
\begin{figure}[t]
\includegraphics [width=8.5cm,height=12cm] {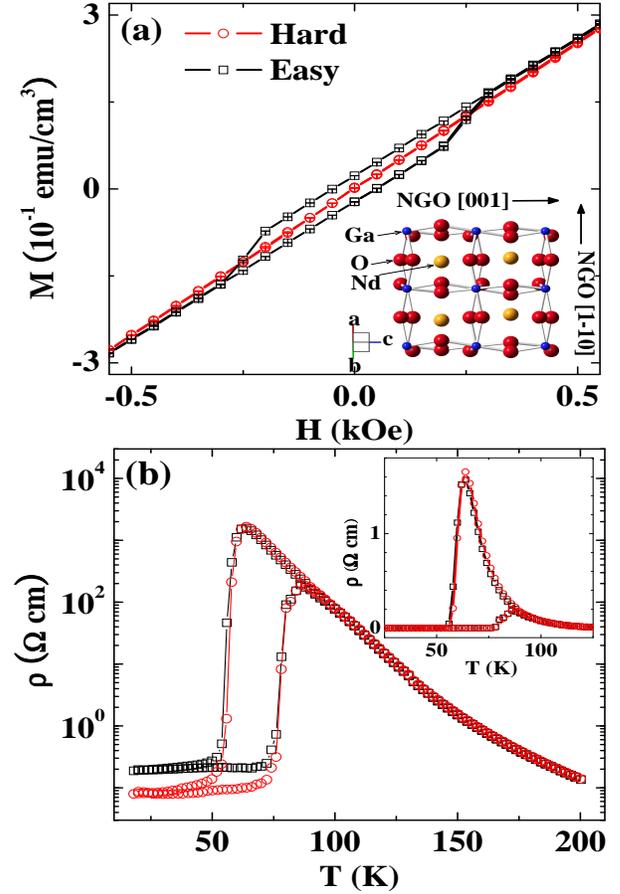}
\caption{(Color online) (a)Magnetic moment (film plus substrate) as a function of field ($M(H)$) of the 30-nm-thick LPCMO thin film grown on NGO with $H$ along hard axis (circle) and easy axis (square) at 60 K. The inset shows atomic structure of NGO ($d_{1\bar{1}0}$ = 3.863 \AA~ and $d_{001}$ = 3.854 \AA). (b) $\rho(T)$ measured along the hard axis (circle) and easy axis (square).}
\end{figure}
    We measured $M(H)$ curves of the LPCMO//NGO films along two perpendicular in-plane directions \textit{viz.}, [1$\bar{1}$0] and [001] directions of the NGO substrates to check for possible strain induced magnetic anisotropy. From Fig. 7(a), it is clear that the [001] NGO direction is the magnetic hard axis, while the [1$\bar{1}$0] NGO direction is the magnetic easy axis of the LPCMO//NGO films. As discussed earlier, when the magnetic field is applied along the easy axis the films show magnetic hysteresis loops below $T_{C}$. Since we measured approximately square shaped thin films (6 mm $\times$ 6 mm) which are chemically in a single phase, we could neglect the in-plane shape anisotropy from the film geometry. Also, when we measured (La$_{1-y}$Pr$_{y}$)$_{1-x}$Ca$_{x}$MnO$_{3}$ ($x$ = 0.33 and $y$ = 0.5) thin films on (110) NGO at 10 K, when it is in fully ferromagnetic state, we still observed an in-plane magnetic anisotropy similar to the anisotropy of ferromagnetic La$_{0.77}$Ca$_{0.33}$MnO$_{3}$ thin films grown on (001) NGO~\cite{RefWorks:94, Dha07}. Thus, we could neglect the shape anisotropy due to elongated or stripe-like FMM regions~\cite{RefWorks:96, RefWorks:102}. Hence, there are two possible interpretations of this magnetic anisotropy \textit{viz.}, magnetocrystalline anisotropy or stress anisotropy~\cite{Cul08}. Murakami \textsl{et al.} suggested that magnetocrystalline anisotropy leads to a domain transition in single crystal (La$_{1-y}$Pr$_{y}$)$_{1-x}$Ca$_{x}$MnO$_{3}$~\cite{RefWorks:96}. Hence, magnetocrystalline anisotropy could also be the reason behind the observed in-plane magnetic anisotropy in our thin films. However, stress anisotropy could also play a critical role due to the different in-plane lattice constants of (110) NGO substrates. Along the [001] NGO direction, the tensile strain on LPCMO is 0.26$\%$ and along the [1$\bar{1}$0] NGO direction the tensile strain is 0.49$\%$ (inset of Fig. 7(a)). The magnetic easy axis was observed to be along the direction with larger tensile strain i.e. the [1$\bar{1}$0] NGO direction (Fig. 7(a)). This result is consistent with experiments on La$_{0.67}$Sr$_{0.33}$MnO$_{3}$ thin films grown under anisotropic tensile stress~\cite{RefWorks:108}. Boschker \textsl{et al.} showed that the in-plane magnetization is due to the compression of the MnO$_6$ octahedra in the out-of-plane direction caused by tensile strain and the uniaxial anisotropy is due to difference in in-plane tensile strains causing further distortion of the octahedra~\cite{RefWorks:108}. Through this mechanism it is possible that if a substrate has different in-plane lattice constants, the magnetic easy axis will be along the direction with higher tensile strain. To distinguish between these two possible origins of magnetic anisotropy, we measured the magnetic anisotropy of LPCMO thin films grown on SLGO substrates which exert negligible stress on LPCMO and there is no in-plane anisotropic stress. From Fig. 4 and Fig. 6(b) it is clear that the magnetic hysteresis loops, $M(T)$, $H_{c}(T)$, and $M_{r}(T)$ of the 30 nm-thick LPCMO//SLGO thin film are almost identical for applied fields along the [100] and [010] directions of the SLGO substrates, at all temperatures. So our observation is that the LPCMO//SLGO film has negligible in-plane magnetic anisotropy, while there is clear magnetic anisotropy in the LPCMO//NGO films. Hence, the main reason for magnetic anisotropy in the LPCMO thin films is the anisotropic stress exerted by the substrate.    
         
    While we observed clear in-plane strain induced magnetic anisotropy, we did not observe any significant in-plane anisotropy in the transport properties (Fig. 7(b)) (we did observe a lower resistivity at low temperatures along the magnetic hard axis but that was due to a higher electric field in that direction since the distance between the voltage leads was shorter along the [001] direction~\cite{Dha07}). In fact, in a linear scale (inset of Fig. 7(b)) the $\rho(T)$ behavior appears identical in the two in-plane directions. The $T_{IM}$ is 60 K in both directions and at that temperature the resistivity anisotropy ($\left|\frac{\rho_{NGO [001]}-\rho_{NGO [1\bar{1}0]}}{\rho_{NGO [1\bar{1}0]}}\right|\times 100$) is $\approx 15\%$. Thus, we believe that anisotropic in-plane properties due to substrate induced strain can be clearly observed in magnetic measurements, but not in transport measurements. Our observation is in contrast to Ward \textsl{et al.} who suggested that the difference in $T_{IM}$ and maximum $\rho$ observed along two perpendicular in-plane directions of LPCMO thin films grown on NGO substrates was due to anisotropic strain and was maximized as magnetic field was lowered~\cite{RefWorks:101}. Such a strain induced resistivity anisotropy has also been predicted theoretically~\cite{Dong10}. It is possible that such resistivity anisotropy can only be clearly observed when the resisitivity measurements are taken at the scale of phase separation i.e. in the micrometer scale, and requires further investigation.\textit{}  
\begin{figure}[t]
\includegraphics [width=8.5cm,height=12cm] {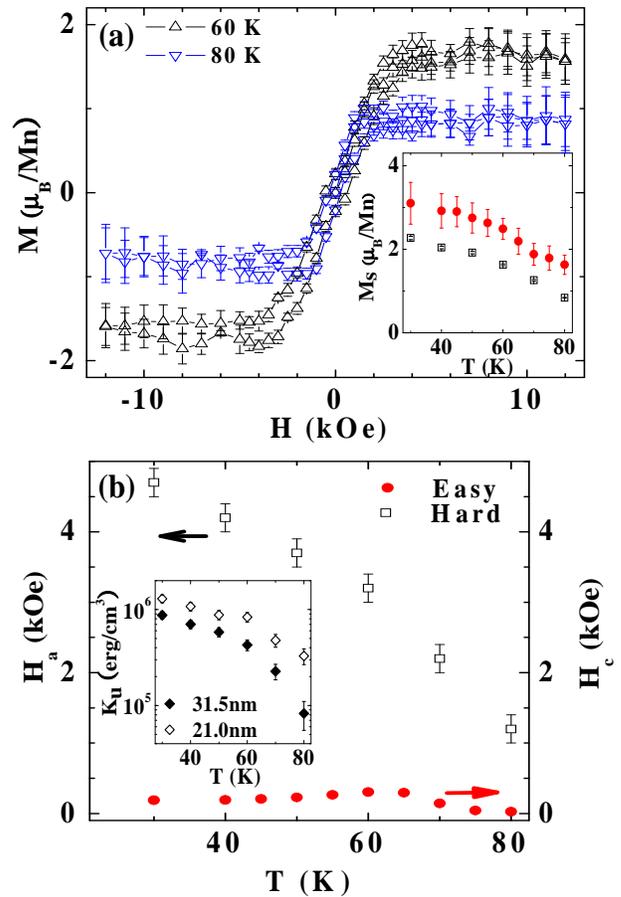}
\caption{(Color online) (a) $M(H)$ curves of the 30-nm-thick LPCMO thin film grown on NGO with $H$ along [001] NGO obtained using linear background fitting. The inset shows saturation magnetization ($M_{s}$) as a function of temperature for the easy axis (circle) and hard axis (square) directions. (b) Anisotropic field ($H_{a}$) (square) and coercive fields ($H_{c}$) (circle) as a function of temperature. The inset shows uniaxial anisotropic constants ($K_{u}$) of the 20-nm-thick (unfilled diamond) and the 30-nm-thick (filled diamond) thin films as a function of temperature.}
\end{figure}
\begin{figure}[t]
\includegraphics [width=8.5cm,height=12cm] {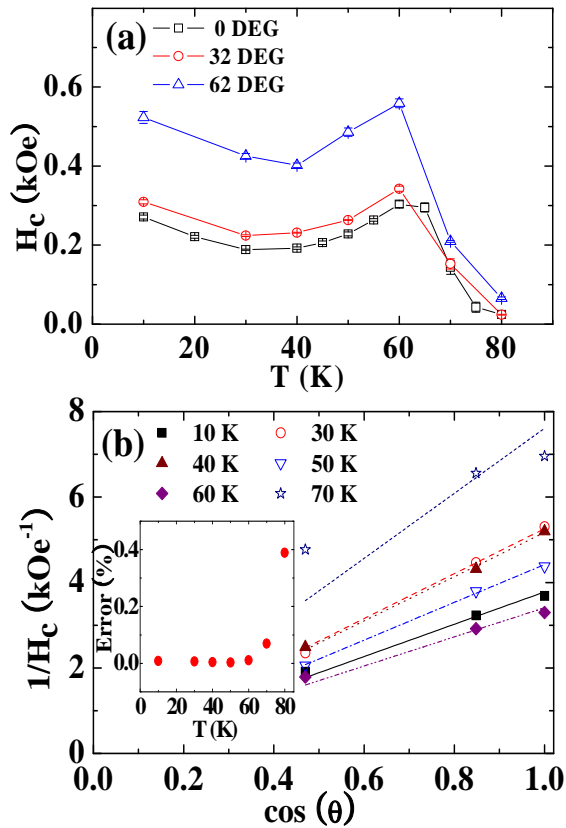}
\caption{(Color online) (a) $H_{c}(T)$ behavior of the 30 nm-thick LPCMO film grown on NGO for three different values of $\theta$, where $\theta$ is the angle between the applied field $H$ and the easy axis. (b) $H^{-1}_{c}$ as a function of cos$(\theta)$ at different temperatures. The inset shows standard error of the slope of the $H^{-1}_{c}$ vs. cos$(\theta)$ graph at each temperature}
\end{figure}

   Next, we ramped the magnetic field up to 12 kOe on the LPCMO//NGO films along the [001] NGO direction i.e. the hard axis. The removal of the background signal was more difficult due to the lack of hysteresis in the $M(H)$ curves. The linear background fitting method was carried out for field values above 5 kOe, since the magnetic moment saturates at a higher field along the hard axis as shown in Fig. 8. Saturation magnetization values along the hard axis direction were slightly lower than those along the easy axis at all the measurement temperatures (inset of Fig. 8(a)). Since the LPCMO//NGO films show strong in-plane uniaxial anisotropy, we could estimate the uniaxial anisotropic constant ($K_{u}$) as a function of temperature in the 20 nm-thick and the 30 nm-thick LPCMO films using saturation magnetization ($M_{s}$) values, anisotropic fields ($H_{a}$) and the relation  $K_{u}$ = $M_{s}$$H_{a}$/2 (inset of Fig. 8(b))~\cite{Cul08}. As expected, the thinner film shows higher $K_{u}$ values due to larger substrate induced strain effects. These values are comparable to that of La$_{0.7}$Ca$_{0.3}$MnO$_{3}$ thin films on NGO substrates (3.6$\times 10^{5}$ erg/cm$^{3}$ at 77 K) and higher than that of a La$_{0.7}$Sr$_{0.3}$MnO$_{3}$ thin film on SrTiO$_{3}$ (8.4$\times 10^{4}$ erg/cm$^{3}$ at room temperature) ~\cite{RefWorks:94,RefWorks:98}. As the temperature was reduced, a gradual monotonic increase of the uniaxial anisotropic constants, saturation magnetization values, and anisotropic fields was observed unlike the non-monotonic behavior of the coercive fields due to the magnetic domain transition. Using the uniaxial anisotropic constant at 60 K and exchange stiffness value ($A$) from Lorentz microscopy, we calculated the critical diameter for domain transition assuming a spherical domain~\cite{RefWorks:96, Kit49}. We obtain a value of 85 nm for the critical radius ($r_{c}$).    
   
   When the anisotropic fields and coercive fields are compared, differences not only in trends from temperature variation but also in the magnitudes can be easily identified (Fig. 8(b)). The anisotropic fields are larger by an order of magnitude or more than the coercive fields along the easy axis. This difference in magnitude implies that Stoner $\&$ Wohlfarth's magnetic reversal mechanism may not be valid~\cite{RefWorks:89, Elb91}. There are two main mechanisms for magnetic switching behavior \textit{viz.}, Stoner $\&$ Wohlfarth's coherent rotation model and a nucleation $\&$ propagation mechanism ~\cite{Cul08}. A direct way to check which model is applicable in our case is a measurement of coercive fields as a function of the angle between the magnetic easy axis of sample and external magnetic fields, since each model predicts a specific behavior. The nucleation $\&$ propagation mechanism leads to the Kondorsky law, ($H_{c}$($\theta$) = $H_{c}$(0)/cos($\theta$))~\cite{Kon0, RefWorks:89, Rei65}. We set [1$\bar{1}$0] NGO (easy axis) as $\theta$ = 0$^{\circ}$ and [001] NGO (hard axis) as $\theta$ = 90$^{\circ}$. $M(H)$ curves were taken at two intermediate angles, $\theta$ = 32$^{\circ} \pm 2^{\circ}$ and $\theta$ = 62$^{\circ} \pm 2^{\circ}$. Fig. 9(a) shows coercive fields as a function of temperature for three different angles. It is clear that the coercive field increases with the angle $\theta$ between the easy axis and the applied magnetic field. When the inverse of coercive fields is plotted as a function of cos$(\theta)$, the Kondorsky law is obeyed between 10 K and 60 K (Fig. 9(b)). This behavior is also observed in other manganite systems grown on artificially miscut substrates~\cite{RefWorks:100}. At the higher temperatures of 70 K and 80K (data not shown for 80 K), the Kondorsky law is not obeyed (inset of Fig. 9(b)). We believe that this deviation from Kondorsky law may be due to the spatial motion of the FMM, AFM-COI, and PMI phases in the FPS state and possible contribution of other reversal mechanisms~\cite{Dha07}.  
   
\section{Conclusions}
   In-plane anisotropic stress leads to uniaxial magnetic anisotropy in manganite thin films. We have shown that when such magnetic anisotropy is induced in phase separated manganites it leads to a single domain to multidomain transition as a function of temperature. The domain transition is similar to that observed in ferromagnetic fine particles as a function of particle size. In phase separated manganites the size of the ferromagnetic metallic regions embedded in a non-ferromagnetic (charge-ordered antiferromagnetic or paramagnetic) matrix increases as the temperature is decreased. Due to the stress induced uniaxial magnetic anisotropy the increase in size of the ferromagnetic regions results in the temperature dependent domain transition. The unique variation of the coercive field with temperature is a signature of the domain transition. The temperature dependence of the coercive field is a feature which could make it possible to use manganites as cryogenic magnetic memory, since magnetic information can be written at a temperature with low coercive field and stored at a lower temperature with higher coercive field.     
   
\section{Acknolwedgements}
This work was supported by NSF DMR-0804452.


\begin{thebibliography}{37}
\bibitem [{afm()}]{afm}%
  \BibitemOpen
  \href@noop {} {\bibinfo  {journal} {For thickness measurement, the thin films
  were partially dipped in the etching solution (H$_{2}$O + KI + 10$\%$ HCl)
  for 10 - 40 seconds. And then, the step, created by etching solution, were
  measured with atomic force microscope.}}\BibitemShut {Stop}%
\bibitem [{tim()}]{tim}%
  \BibitemOpen
  \href@noop {} {\bibinfo  {journal} {T$_{IM}$ and T$_{MI}$ were defined as the
  maximum or minmum change of resistivity as a function of
  temperature.}}\BibitemShut {Stop}%
\bibitem [{zfc()}]{zfc}%
  \BibitemOpen
  \href@noop {} {\bibinfo  {journal} {With direct subtraction, the zero field
  cooled (ZFC) warming mangetization measurement showed a negative at low
  temperature. However, when the LPCMO was at near to Ferromagnetic to
  Paramagnetic transition, the rise and decrease of magnetic moments were
  observed at the same temperature.}}\BibitemShut {Stop}%
\end{thebibliography}
\end{document}